# Physics-informed machine learning applied to the identification of high-pressure elusive phases from spatially resolved X-ray diffraction large datasets

Lucas H. Francisco,[1,2]*† Camila M. Araújo,[1] André A. M. C. Silva,[1,2] Ulisses F. Kaneko,[1]§ Jairo Fonseca Jr,[1] Guilherme A. Calligaris,[1] Audrey D. Grockowiak,[1]‡ Danusa do Carmo,[1] Ricardo D. dos Reis,[1] Narcizo M. Souza-Neto[1]**

1 Brazilian Synchrotron Light Laboratory (LNLS/SIRIUS), Brazilian Center for Research in Energy and Materials (CNPEM), Campinas 13083-970, São Paulo, Brazil
2 Instituto de Física Gleb Wataghin, Universidade Estadual de Campinas, Campinas, SP, 13083-859, Brazil
*Corresponding author. Email: lucas.francisco@cnpem.br
**Corresponding author. Email: narcizo.souza@lnls.br
† Present address: Brazilian Center for Research in Energy and Materials, Campinas 13083-970, São Paulo, Brazil
‡ Present address: Leibniz Institute for Solid State and Materials Research, IFW Dresden, 01069 Dresden, Germany
§ Present address: São Paulo State University (Unesp), Institute of Geosciences and Exact Sciences, Rio Claro

## Abstract

Multi-technique high resolution X-ray mapping enhanced by the recent advent of 4th generation synchrotron facilities can produce colossal datasets, challenging traditional analysis methods. Such difficulty is clearly materialized when probing crystal structure of inhomogeneous samples, where the number of diffraction patterns quickly increases with map resolution, making the identification of crystal phases within a vast collection of reflections unfeasibly challenging by direct human inspection. Here we develop a novel analysis approach based on unsupervised clustering algorithms for identifying independent phases within a diffraction spatial map, which allowed us to identify the material distribution across a high-pressure cerium hydride. By investigating the specific compound, we also contribute to the understanding of synthesis inhomogeneities among the superhydrides, a prominent superconductor class in condensed matter physics whose characterization is highly challenging even for state-of-the-art materials techniques. The analysis framework we present may be readily extended to any correlated set of curves whose features are tied to specific entities, such as structural phases.

## Teaser

A novel clustering-based approach for analyzing big, correlated curve sets is applied to X-ray diffraction 2d mapping of hydrides.

## Introduction

The need to determine material properties, such as crystal and electronic structures, has driven the development of state-of-the-art techniques and instruments over the past several decades. Today, advanced 4$^{th}$ generation synchrotron sources with higher radiation brilliance [1,2] combined with extensive data produced by cutting-edge X-ray detectors [3–6] may generate much more data, beyond the capability of a human to analyze it individually. This is especially true when studying spatially resolved crystal structures over a sample volume, in which case each pixel or voxel might accommodate a rich set of information. One example is the recent quest for high temperature superconductivity in the superhydrides class of materials [7–9], which needs to be synthesized under extreme conditions that may intrinsically induce non-homogeneous phases, thus requiring spatially resolved analysis. To this end, one wants to employ X-ray techniques able to identify crystal structures within a sample, such as X-ray diffraction (XRD).

The standard methods for interpreting diffraction data directly employ crystal models and analysis techniques based in the modern crystallography toolset [10,11], and have been tremendously successful finding applications in diverse areas, ranging from metallurgy, geology and semiconductors to biology and pharmaceuticals. However, as the amount of data produced in modern facilities increases significantly, the analysis of individual diffraction patterns becomes impractical. Efforts are needed in optimizing data acquisition and analysis processes to improve scientific reproducibility, transparency, and reliability, opening paths to new discoveries. One possibility is to use Artificial Intelligence (AI) methods to speed up and expand the robustness of data analysis methods. AI strategies already find broad, though not necessarily routine, applications in solid state science, such as physical properties modeling for materials discovery [12,13] and big data analysis, including X-ray diffraction investigations [14], in which unsupervised learning clustering algorithms have been applied through diverse approaches and varying extents [15–20]. Such techniques are posed to play an essential role in extensive datasets analysis undertaken in large-scale user facilities. Here we present a novel and flexible workflow strategy for applying AI clustering algorithms to large feature-rich correlated curves datasets, using X-ray diffraction mapping of high-pressure hydrides as our model problem. In our application case, the technique allows us to automatically identify unique independent crystal phases associated with angular regions of the obtained diffractograms. Though straightforward, this approach is powerful and highly promising in elevating our capability to process and identify unique crystal phases from limitless datasets and determine meaningful features for multimodal diffraction imaging [21], while its core idea is flexible and applicable to many other multidimensional curve set cases.

The materials class to which we applied this method, the novel high $T_c$ superconducting hydrides [7–9], displays the most significant challenges in determining spatially resolved crystalline phases today. These materials are often synthesized at high pressures and temperatures, in the Mbar range and above 1000 – 2000 K, requiring the use of laser-heated diamond anvil cells [9,22]. Such a method is known to result in significant inhomogeneity in both pressure and temperature [23–26], which may often lead to the formation of different phases within the experimental pressure chamber, resulting in non-homogeneous samples. This scenario makes it challenging to obtain accurate structural information using traditional X-ray diffraction methods making the superhydrides an ideal case to apply our new approach. The advances in spatially resolved 2D X-ray diffraction using nanometer- and micrometer-sized X-ray beams allowed us to obtain detailed structural information maps from these materials, including sample inhomogeneity, providing us the dataset we utilize here to apply and demonstrate the approach and identify unique crystal phases.

Within the example use case of superhydride synthesis, we selected the cerium superhydride as our benchmark material. The experimental attainment of cerium superhydrides was first reported by Li et al. in 2019 [27], followed by a report by Salke et al.[28] in the same year, while Peng et al. [29] had already theoretically predicted stable structures of $CeH_n$ for n = 3, 4, 8, 9, 10 in 2017. Li et al. performed a high-pressure reaction, at 135 GPa, of Ce and $H_2$, without heating, and observed phases with compositions $CeH_3$, $CeH_{3+x}$, $CeH_4$, $CeH_{9-x}$, and $P6_3/mmc$ $CeH_9$. Salke et al. [28] also successfully synthesized $P6_3/mmc$ $CeH_9$ (in addition to $CeH_3$ $Pm3/n$ and $CeH_2$ $Fm3/m$ at lower pressures) by using laser heated diamond anvil cells at 80 – 100 GPa and temperatures up to ~1700 K and reported calculations predicting that the $CeH_9$ phase would exhibit a superconducting $T_c$ of 105 – 117 K at 200 GPa [28], making it a high-temperature superconducting superhydride. The $CeH_9$ superconducting transition was measured in 2021 by Chen et al. [30] with transition temperatures between 49 K and 123 K and pressures from 88 GPa to 165 GPa in different cells, inside which the superconducting phase was synthetized through a laser heated reaction between metallic Ce and ammonia borane ($NH_3BH_3$). Literature findings thus indicate the robustness of $CeH_9$ synthesis at a relatively low-pressure regime (~100 GPa) compared to other superhydrides, making it a good benchmark ground for the application of our clustering-based methodology on superhydrides and heterogeneous samples in general.

# Results

A laser-heated diamond anvil cell (DAC), subsequently confirmed to contain cerium superhydrides, was prepared as described in the Methods section. We present the pressure chamber contents at 124 GPa after synthesis on Figure 1a, in which a solid cerium-based sample is centered in the circular hole of the rhenium gasket enclosing the chamber region. To investigate the multiple phases synthesized inside the DAC, we explored the Gaussian micro-focused (2×2 microns) X-ray beam of the Sirius EMA beamline [31] to perform a two-dimensional X-ray diffraction map of the sample using 25 keV radiation (0.495937 Å). Figure 1b presents the X-ray beam intensity transmitted by the cell as it is shifted along the transversal x, y direction (perpendicular to the longitudinal beam direction), where the gasket hole leads to a plateau used for determining the central $x_c$ and $y_c$ coordinates at the center of the pressure chamber, 12.4290 mm and 50.0915 mm on the beamline motors coordinate system, both defined as 0 mm in the gasket center coordinate system. The DAC movement allowed us to map a 10 × 11 square pattern 33 μm tall around the ($x_c$, $y_c$) position, covering most of the sample space of the diamond anvil cell, resulting in a diffraction map with a 3.3 μm pixel size, as shown in Figure 1c, which also shows selected diffractograms with distinct features from different mapped points. Points may be labeled by integer indices (i, j), with i ranging from –4 to +4 and j ranging from –5 to +5, plus the point (–5, –5) at the top left on the map.

**Semi-automatic pre-processing**

The large dataset analysis procedure comprises a semi-automatic pre-processing part followed by an automatic part based on hierarchical clustering. A complete analysis script, including both analysis parts, may be found in Data S1. The semi-automatic part starts with the diffractograms normalization, in which all curves were divided by the incident X-ray intensity (in our case, we used the synchrotron storage ring current mean value during each x-ray measurement). After normalization, we performed a background removal step. Various methods might be used for this step to reach a flat baseline which does not vary among diffractograms and therefore does not introduce spurious variation at localized 2θ intervals. For our data, we modeled the background as a polynomial function fit to points of specific fixed positions in the diffractograms, chosen so that no diffraction peaks appeared at their positions. The pre-processing step leads to a diffractograms set linked to the mapped points, as displayed in Figure 1d.

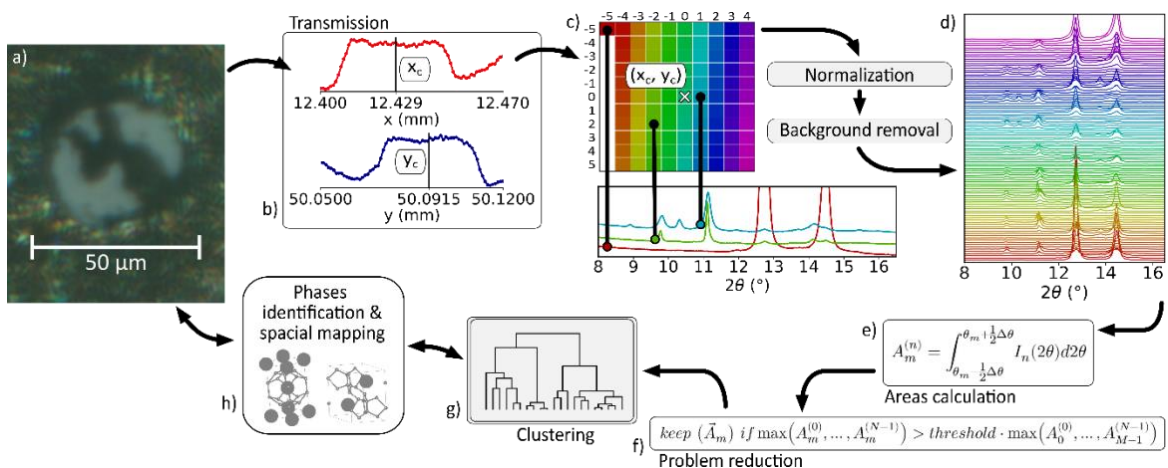

**Fig. 1. Pressure chamber and analysis procedure overview.** (**A**) image of the rhenium gasket hole in which a solid cerium-based sample is surrounded by a pressure medium; (**B**) centering procedure using X-ray intensity transmission through DAC, in which the defined $x_c$ and $y_c$ center coordinates are indicated; (**C**)

square grid centered on ($x_c$, $y_c$) at which diffractograms were collected, as shown for selected points; **(D)** normalized diffractogram patterns obtained along the gasket hole after normalization and background removal; **(E)** area vector components as defined from the obtained diffractograms; **(F)** problem complexity reduction by eliminating area vectors which do not reach significant values anywhere on the map when compared to the maximum observed intensity throughout the dataset by a given scaling threshold; **(G)** and **(H)** are the final steps on the procedure, the hierarchical clustering algorithm application to the areas (the core of our approach) and the analysis of clustering results, allowing for easy phase mapping and identification.

**Unsupervised clustering of diffraction data**

The core of our approach is the analysis of the partitioning of all the diffractograms in the 2D map and the grouping of their intensities in a series of area vectors. For describing such partitioning, we consider the set of $N$ diffractograms on a map, indexed by $n = 0, ..., N - 1$ and labeled $I_n(2\theta)$ (intensity of the n-th diffractogram as a function of the angle $2\theta$). Then, we describe the partitioning by $M$ intervals in $2\theta$, labeled by $m = 0, ..., M - 1$, of same width $\Delta\theta$ and centered in positions $\theta_m$. We chose a 0.05° partitioning length ($\Delta\theta$) for our dataset, which is smaller than the observed peak widths of tenths of degrees, thus minimizing partitions positioning effects, while being larger than the 0.0042° step size at the analyzed curves, thus minimizing the effect of high frequency noise. Then we convert the diffractograms information to a set of area vectors $\vec{A}_m$ ($m$ from 0 to $M - 1$), step on Figure 1e, whose components are the integrated intensities of a given $[\theta_m - (1/2)\Delta\theta, \theta_m + (1/2)\Delta\theta]$ interval at the various $I_n(2\theta)$ diffractograms:

$$\vec{A}_m = \begin{pmatrix} A_m^{(0)} \\ \vdots \\ A_m^{(n)} \\ \vdots \\ A_m^{(N-1)} \end{pmatrix} \quad (1)$$

$$A_m^{(n)} = \int_{\theta_m - \frac{1}{2}\Delta\theta}^{\theta_m + \frac{1}{2}\Delta\theta} I_n(2\theta) d2\theta \quad (2)$$

Each area vector thus represents the behavior of a given diffractogram region over the map and we may define a correlation $corr(\vec{A}_{m'}, \vec{A}_{m''})$ between any pair of area vectors, as illustrated in Figure 2a, leading to an $M \times M$ correlation matrix. Our approach is then based on the fact that vector areas pair in $2\theta$ regions whose intensities across the map are dominated by the diffraction from the same phase are expected to present strong positive correlations. In contrast, area pairs in regions with different phases show low correlation. Before calculating the correlations matrix, an additional step is performed, as identified in Figure 1f, and it consists of eliminating $2\theta$ regions with low diffraction intensity throughout the entire map. This is done by finding the maximum $A_m^{(n)}$ value across the entire dataset and eliminating any area vector $\vec{A}_m$ whose largest component is smaller than a chosen threshold percentage of such $A_m^{(n)}$. Here, we chose a 0.2% threshold, eliminating $2\theta$ regions which never get bigger than 0.2% of the maximum area.

After the complexity reduction step, the correlation matrix is calculated based on the area vectors normalized to unit norm, presented in Figure 2b, in which both axes are labeled by area indices $m$. A 1.0 value represents complete correlation, in which the intensity of the components for a given vector is directly proportional to the component intensities of the other vector, which is ideally expected for regions with the pure presence of diffraction from a single phase and trivially occurs for the main matrix diagonal. Correlations of -1.0 represent the opposite behavior, where the components for one area increase when those of another area decrease. In null correlation, the components for one area do not show dependence on the components of the other.

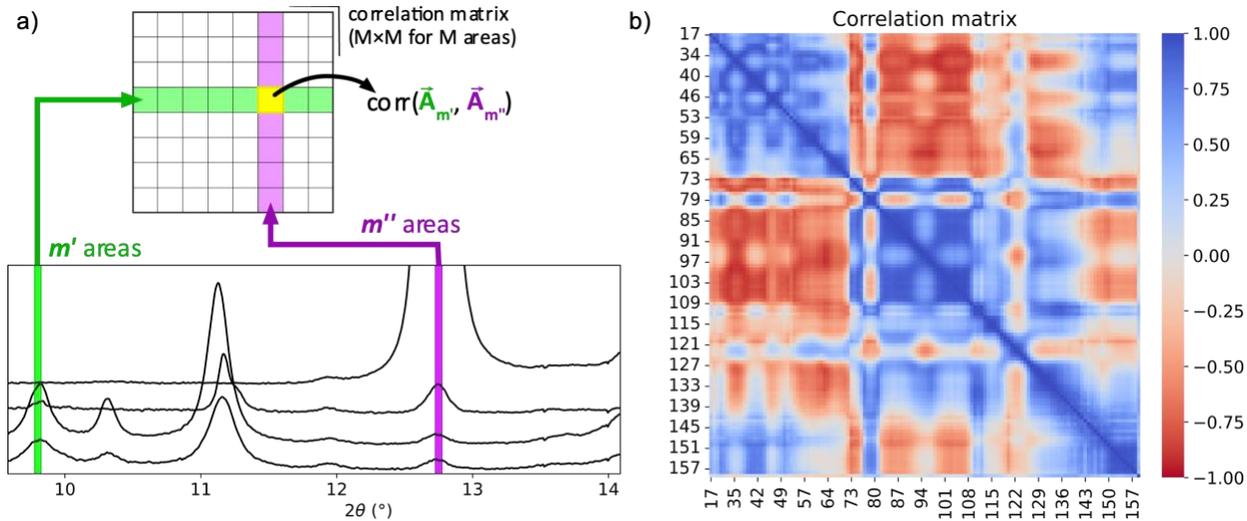

**Fig. 2. Correlation matrix obtained from diffractograms set.** (A) illustration of correlation matrix subset in which the element relating areas m' and m'' is highlighted; (B) correlation matrix between area vectors normalized to unit norm from our dataset, whose indices are present in the two Cartesian axes (correlation between components). The resulting matrix is symmetric since the two vectors' correlation is reciprocal.

The matrix allows us to identify $2\theta$ regions with a greater probability of belonging to the same crystalline phase as regions of positive correlation outside the main diagonal. Based on this observation, a clustering algorithm may be applied to the problem to group all the area vectors concerning their correlation, resulting in clusters of grouped areas and, therefore, $2\theta$ regions identifiable with a single crystalline phase (a single set of highly correlated diffraction peaks). Here, we have chosen the hierarchical clustering algorithm [32], implemented in the SciPy package [33], with a distance metric between normalized diffraction areas $\vec{A}_{m'}$ and $\vec{A}_{m''}$, being defined from their $r_{m'm''}$ correlation as:

$$d(\vec{A}_{m'}, \vec{A}_{m''}) = 1 - r_{m'm''} = 1 - \frac{(\vec{A}_{m'} - \langle\vec{A}_{m'}\rangle) \cdot (\vec{A}_{m''} - \langle\vec{A}_{m''}\rangle)}{|\vec{A}_{m'} - \langle\vec{A}_{m'}\rangle||\vec{A}_{m''} - \langle\vec{A}_{m''}\rangle|} \quad (3)$$

In which $\langle\vec{A}_m\rangle$ represents the average $\vec{A}_m$ vector and $|\vec{A}_m|$ its cartesian length. Such metric leads to zero distance between an area and itself and highest distance for the negatively correlated areas. We also chose a complete linkage method for the clustering, which defines the distance between clusters as the highest distance between any pair of elements, one in each cluster, a choice that guarantees that highly uncorrelated areas do not get inappropriately grouped. This clustering step is done automatically and unsupervised, it is the final step of our procedure as presented in Figure 1g and its results allow for easy phase mapping and identification (Figure 1h).

**Hierarchical Clustering Results**

The results of the hierarchical clustering procedure are a hierarchy of clusters, as depicted in Figure 3a, denominated dendrogram, which represents the entire clustering sequence, and in which base each vertical line, also called a leaf, corresponds to a diffractogram area integrated around a specific $\theta_m$ center value, as illustrated by the inset in Figure 3a. The diagram also represents and illustrates the algorithm workings, which runs in iterative steps, starting from the single element clusters (singleton clusters, shown as the diagram leaves) where, in each step, clusters with the lowest distances (defined by metric and linkage method) merge in a single cluster, up to the final step, in which all the elements merge into a single cluster. Each cluster merge is represented in the diagram as a horizontal line, whose height is the distance between the merged clusters.

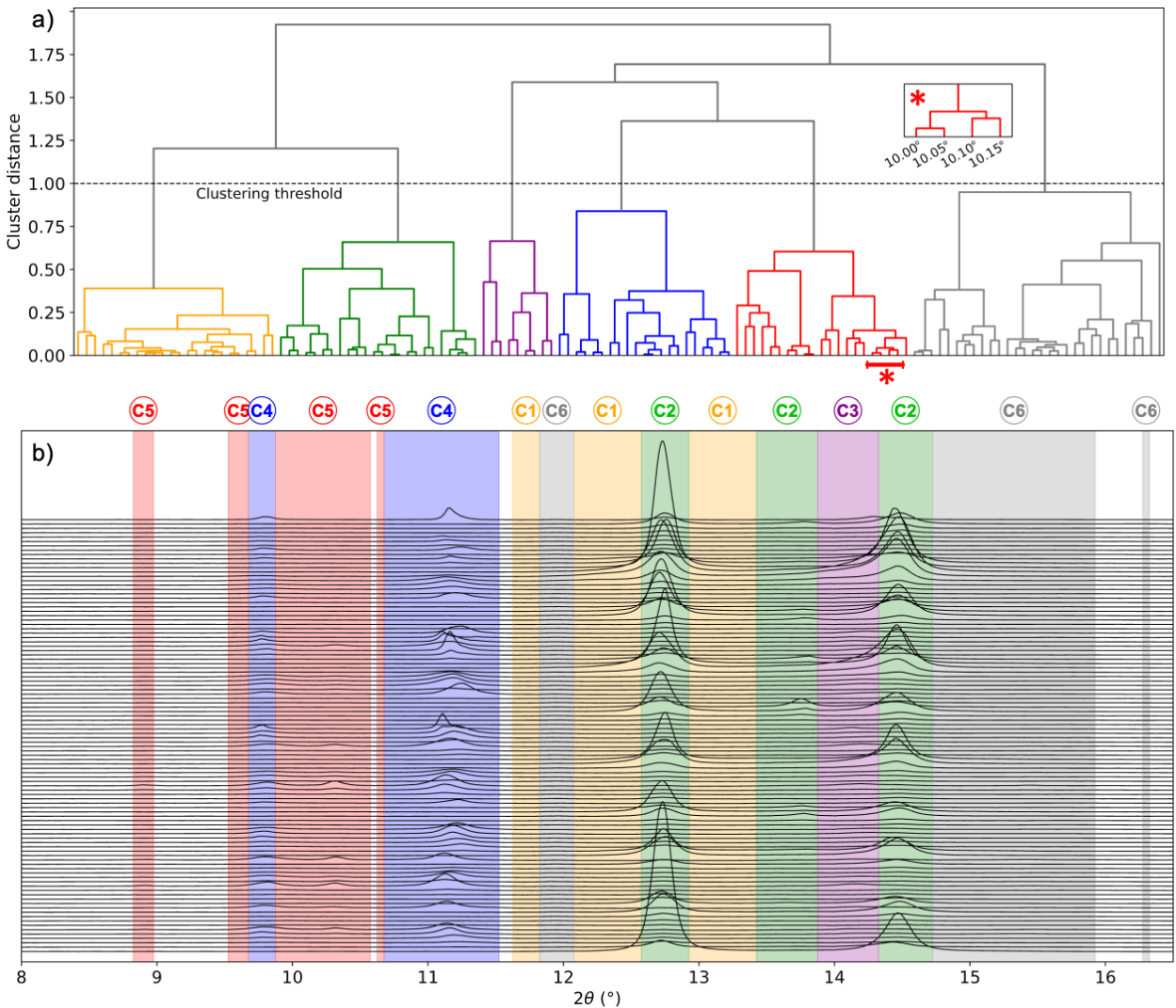

**Fig. 3. Hierarchical clustering results.** (**A**) Dendrogram showing hierarchical clustering to identify independent phase regions. Vertical lines are elements, which might be clusters of areas or single areas (singleton clusters). Horizontal lines represent clustering steps; their height is the distance between the two merged clusters. Inset exemplifies the beginning of a clustering, where areas centered at the angular positions on the horizontal axis are joined in the indicated sequence, eventually inserted in the 5[th] cluster. (**B**) Diffractograms showing the area intervals associated with the clusters; clusters C1 to C6 run from left to right on the dendrogram with color correspondence to the diffractograms shaded areas.

From the resulting dendrogram for our dataset we have chosen 6 clusters indicated by different colors and labeled from left to right on Figure 3a, and the corresponding diffractogram regions are shown in Figure 3b. We chose the number of clusters based on the vertical distances in the diagram, where the union of the 6 clusters begins with a minimum length of 1.20 (negative correlation of -0.20 on the union between the 1st and 2nd clusters), while the last union from the 6 clusters is performed at a length of 0.95 (positive correlation of +0.05). Therefore, the classification into 6 clusters unites areas with positive correlation without connecting them to zero or negative correlation areas. The complete linkage method guarantees this since it considers the most significant distances between elements of the clusters.

**Crystal Phases Identification**

The clustering process concluded with identifying independent crystallographic peak sets associated with the dendrogram leaves of the 6 clusters, from which C2, C4 and C5 were the main clusters containing substantial peaks (Figure 3b). We then performed peak assignment to crystal phases by using literature information and calculations for expected peak angles and intensities at the measurement pressure (124 GPa) using the VESTA [34] software. Cluster 2 peaks may be promptly assigned to rhenium reflections from the gasket, as main reflections around 12.8° and 14.6° and a less intense reflection around 13.8° are expected at 124 GPa, as calculated from the lattice given by an equation of state [35]. We were then able to associate the peaks at clusters C4 and C5 to the reflections corresponding to the proposed experimental Ce superhydride structural phases as reported by Li et al. [27]. Cluster C5 peaks were associated with the reflections (100) and (101) from the $P6_3/mmc$ CeH$_9$ phase, and cluster C4 peaks were ascribed to the reflections (002) and (101) from the $I4/mmm$ CeH$_4$. The cluster–phase association is summarized in Figure 4, which contains integrated intensity maps of each full cluster region and a combined map created by assigning each crystal phase to an RGB channel.

For both assigned phases, the reflection positions (Figure 4a) were enough for completely determining the lattice parameters. The results for cell volume and c/a ratio for the obtained hexagonal CeH$_9$ are 64.57 Å$^3$ and 1.490, which may be compared with the literature values 63.241 Å$^3$ and 1.5456 by Li et al at 159 GPa [27] and 62.836 Å$^3$ and 1.4936 by Salke et al at 100 GPa [28]. For the tetragonal CeH$_4$ phase, our peak positions lead to 47.10 Å$^3$ and 2.04, which may be compared to the literature values 46.046 Å$^3$ and 1.9747 by Li et al at 76 GPa [27]. The clustering method did not identify the reflections (110) and (002) of CeH$_9$ and (110) of CeH$_4$, which is conceivably due to the presence of intense gasket peaks dominating the regions and thus the area vectors, which were classified in a single cluster, referring to the rhenium gasket peaks (identified in the clusters C1, C2 and C3). Cluster C6 includes lower intensity regions, which could be related to pressure medium material or artifacts from differences in beam transmission.

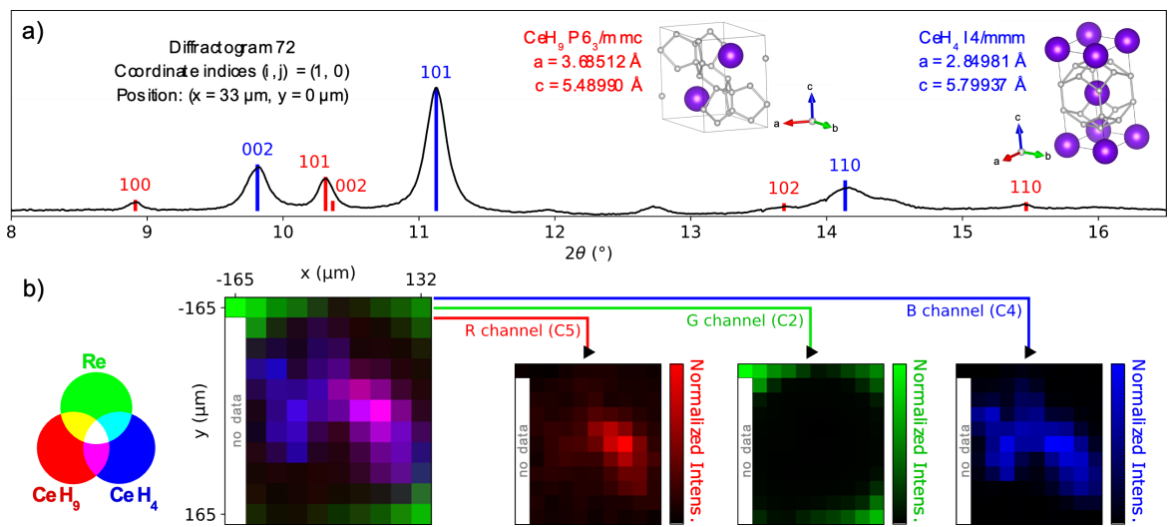

**Fig. 4. Phase map assignments.** (A) X-ray diffraction pattern obtained for point with coordinate indices (1; 0), displaying high-pressure hydrides at 124 GPa and the peaks calculated from the phases included in the figure; (B) X-ray diffraction maps composed of the normalized intensities of clusters C2, C4, and C5 associated with, respectively, green, blue, and red channels, and the phases of Rhenium, $CeH_4$, and $CeH_9$. The combined RGB map provides a reconstruction of the structural phases distribution spatial map based on the clustering method results.

## Discussion

The crystal phases identification step performed for our model case revealed a great benefit from our usage of peak clustering: since groups of correlated peaks were separated, when ascribing them to specific phases we are only concerned with the peaks of one specific cluster at a time. No matter how many peaks there are in a specific diffractogram, we expect that the ones outside the currently analyzed cluster are unimportant to the currently sought phase. Also, during this process, one already knows the spatial distribution of the phase being determined, given by the cluster intensity map, such as the ones in Figure 4b. The diffraction angle ranges from independent phases and their intensity maps, given by the unsupervised clustering, are much more easily identifiable as crystal phases when compared to the information from a large number of diffractograms with many varying correlated peaks.

Our approach thus allowed us to directly identify the structural phases within the sample synthesized under extreme pressures and temperatures, obtaining their spatial distribution. We were then able to assess these inhomogeneous phases with the X-ray diffraction data from the multimodal map. While the $CeH_4$ phase seems easily obtained and abundant, the $CeH_9$ phase, the high-$T_c$ superhydride, is found to be spatially restricted to a region most likely to have contained the optimum synthesis condition, as observed in Figure 4b. This result stresses the importance of evaluating the inhomogeneous nature of a sample synthetized under extreme inhomogeneous conditions to properly understand other position-dependent physical quantities eventually being measured, such as transport properties.

Such conclusion is a direct contribution to the superconducting superhydrides and extreme conditions synthesis research in general. Yet, it is key to emphasize that out approach can be generally applied to any large set of feature-rich correlated curves, and, more generally, leads to an unsupervised procedure to automatically process large datasets such as the ones produced in 4$^{th}$ generation synchrotron beamlines, for which the solely human-based analysis is insufficient. Such as in our application case, the high-resolution information thus achievable allows for the assessment of hidden phases within small portions of inhomogeneous materials. This prospect paves the way for novel materials identification using nanometer resolution maps when, for example, new crystalline phases are sought within inherently inhomogeneous samples synthesized at the most diverse thermodynamic conditions, including possibly even more extreme ones.

This combination of automated data analysis of large structural datasets with extreme conditions synthesis of materials under non-conventional environments and cutting-edge probing methods may therefore open new doors for materials discovery.

## Materials and Methods

To synthesize $CeH_x$, we used a diamond anvil cell (DAC) assembled with single beveled diamonds of 100 µm culets fastened to the cell using graphite adhesive. We used a rhenium gasket, which was pre-indented at the experimental DAC to a minimum thickness of ~20 µm, at which a 35 µm diameter hole was drilled by electrical discharge machining (EDM) to form the pressure chamber. The synthesis was performed similarly to the one used by Chen et al. [30]. The DAC was loaded with ammonia borane ($NH_3BH_3$, 97% Sigma-Aldrich) as pressure medium and hydrogen source, and with a metallic piece of cerium of dimensions close to the gasket hole. The loading geometry was so that the cerium piece was in direct contact with the diamond at the side of the DAC opposite to the one receiving the laser heating radiation. The loaded cerium was obtained from a larger 99.8% nominal purity piece which had been previously abraded to expose non-oxidized material and the loading was performed inside a glove box ($pO_2$ < 10 ppm and $pH_2O$ < 0.1 ppm) to avoid exposing the ammonia borane to atmospheric humidity. We then submitted the material to pressures up to 127 GPa and heating up to 2300 K by laser heating with infrared radiation (1064 nm) modulated at 5 kHz and focused (spot size ~10 $\mu m$) at the center of the cerium metallic piece. The pressure was measured using the culet Raman diamond edge [36], excited by 532 nm radiation, and a pressure of 124 GPa was observed at the final DAC state. Figure 1a shows the experimental pressure chamber (gasket hole) after synthesis, when the gasket hole reached a diameter of about 29 µm.

As presented at the Results section, the micro-focused 2×2 microns X-ray beam of the Sirius EMA beamline [31] was used to perform a two-dimensional X-ray diffraction mapping of the sample using 25 keV radiation (0.495937 Å). The pressure cell was mounted at the beamline in a sample holder connected to a positioning hexapod. The observed x-ray transmission was used to relate the sample holder absolute position to a local coordinate system centered at the gasket hole, as shown in Figure 1b. Such coordinate system, used for referring to the measured points, was properly transformed to ensure visual and geometrical comparison between the crystal phases map in Figure 4b and the microscopy image in Figure 1a. The diffraction patterns were collected with 300 s integration time using an area detector whose distance to the sample had been previously calibrated using a standard $LaB_6$ loaded dummy DAC, positioned at the same sample holder. Such calibration and the azimuthal integration of diffraction patterns were performed using the DIOPTAS software [37], The procedure allowed us to measure the diffractograms in the square pattern presented in Figure 1c, with a 3.3 µm pixel size given by the spacing between data collection positionings of the DAC, covering most of the sample space.

## Acknowledgments

We thank Flávio Garcia for stimulating discussions and helping during the XRD data collection. This research used facilities of the Brazilian Synchrotron Light Laboratory (LNLS), part of the Brazilian Center for Research in Energy and Materials (CNPEM), a private non-profit organization under the supervision of the Brazilian Ministry for Science, Technology, and Innovations (MCTI). The EMA beamline and LCTE group staff are acknowledged for the assistance in the DAC preparations and high-pressure experiments during proposals 20210092 and 20210145.

**Funding:**

Serrapilheira Institute grant Serra-1709-17301 (NMSN)
Fundação de Amparo à Pesquisa do Estado de São Paulo grant 2018/10585-0 (LHF)
Fundação de Amparo à Pesquisa do Estado de São Paulo grant 2018/00823-0 (RDR)
Fundação de Amparo à Pesquisa do Estado de São Paulo grant 2013/22436-5 (NMSN)

**Competing interests:**

All authors declare they have no competing interests.

## Data and materials availability:

All data is available in the main text or the supplementary materials.

## Supplementary Materials

**Supplementary materials include:**

**Other Supplementary Materials for this manuscript include the following:**

Data S1: Compressed file containing all raw data processed in the main text and code for performing a hierarchical clustering-based analysis leading to the cluster maps for cerium superhydrides and rhenium at the diamond anvil cell.

**Supplementary Text**

Data S1 (separate file, may be requested to the authors) contains all XRD data necessary for running the analysis presented in the main text. It also contains Python code for performing a complete hierarchical clustering-based analysis on such data leading to the 2D cluster maps.

The analysis sample code has been tested in Python version 3.12.

Data and codes are separated in the following structure inside a compressed file:
- *requirements.txt* – requirements file for setting up Python environment.
- *data/*.xy* – diffractogram files used in example analysis, corresponding to the XRD data presented in the main text for a cerium-based sample inside a diamond anvil cell.
- *data/Sirius_mA_2022-01-09-06-36-47_18h.csv* – Sirius storage ring current during the experiment, used for diffractogram normalization in the analysis.
- *example-analysis.ipynb* – main analysis file that might be followed to perform a complete clustering-based procedure for determining diffractogram maps from the diffraction data. Depends on the separate code files described below.
- *diffractogram.py* – file containing custom class for representing a diffractogram in the map.
- *xrdmap.py* – file containing custom class for representing a diffractogram map.
- *utils.py* – utility functions file, mainly aiming curve analysis.

**Data S1. (separate file)**

Compressed file containing all raw data processed in the main text and code for performing a hierarchical clustering-based analysis leading to the cluster maps for cerium superhydrides and rhenium at the diamond anvil cell.